\documentclass[doublecol]{epl2}
\usepackage{amssymb}
\usepackage{amsmath}
\usepackage{bbm}
\usepackage{graphicx}
\usepackage{color}

\begin{document}
\author{Andreas Herrmann \inst{1} \and Yuta Murakami \inst{1} \and Martin Eckstein \inst{2} \and Philipp Werner \inst{1}}

\institute{                    
  \inst{1} Department of Physics, University of Fribourg, 1700 Fribourg, Switzerland\\
  \inst{2} Department of Physics, University of Erlangen-N\"urnberg, 91058 Erlangen, Germany %\\
}
\title{Floquet prethermalization in the resonantly driven Hubbard model}

\date{\today}

\hyphenation{}

\abstract{
We demonstrate the existence of long-lived prethermalized states in the Mott insulating Hubbard model driven by periodic electric fields. These states, which also exist in the resonantly driven case with a large density of photo-induced doublons and holons, are characterized by a nonzero current and 
an effective temperature of the doublons and holons which depends sensitively on the driving condition. 
Focusing on the specific case of resonantly driven models whose effective time-independent Hamiltonian in the high-frequency driving limit corresponds 
to noninteracting fermions, we show that the time evolution of the double occupation can be reproduced by the effective Hamiltonian, and that the prethermalization plateaus at finite driving frequency are controlled by the next-to-leading order correction in the high-frequency expansion of the effective Hamiltonian.        
We propose a numerical procedure to determine an effective Hubbard interaction 
that mimics the correlation effects induced by 
these higher order terms. 
}

%\pacs{71.30.+h}{Metal-insulator transitions and other electronic transitions}

\maketitle

{\it Introduction.} The properties of materials can be tuned by chemical substitution, applied pressure, or static external fields. In the theoretical description, these modifications correspond to changes in the Hamiltonian parameters or the addition of extra terms describing the applied fields. In recent years, ``Floquet engineering" has emerged as a versatile tool which enables new levels of control \cite{Bukov2015}. The idea is to apply periodic perturbations to a system which lead to modified parameters or new terms in the effective static Hamiltonian describing the ``stroboscopic" evolution from one period to the next. The theoretically predicted phenomena range from hopping renormalizations \cite{Eckardt2005,Tsuji2011} and modified exchange couplings \cite{Mentink2015,Kitamura2016,Eckstein2017} to synthetic gauge fields \cite{Bukov2015,Bukov2016} and topological phase transitions \cite{Oka2009,Lindner2011,Kitagawa2011}. Some of these effects have been demonstrated in cold atom systems \cite{Struck2012,Aidelsburger2013,Jotzu2014,Goerg2017} and laser-irradiated topological insulators \cite{Wang2013}. 

Since the Hamiltonian of a periodically driven system satisfies $H(t+T)=H(t)$, where $T$ is the period, the time evolution operator $U(t_2,t_1)$ can be written as $U(t_2,t_1)=e^{-iK_\text{eff}(t_2)}e^{-iH_\text{eff}(t_2-t_1)}e^{iK_\text{eff}(t_1)}$, where $H_\text{eff}$ is a time-independent static Hamiltonian and $K_\text{eff}(t)$ the so-called kick operator \cite{Bukov2015}. 
In the regime of large driving frequency $\Omega=\frac{2\pi}{T}$ one can derive $H_\text{eff}$ by a high-frequency expansion,  which is truncated at a given order in $\frac{1}{\Omega}$ \cite{Bukov2015,Mori2016,Mikami2016}. 
While the resulting effective model may contain new terms associated with interesting physical effects or novel phases, 
low-energy properties of this model 
can only be realized if 
heating effects are small, and the driven system is stuck in a long-lived quasi-steady state exhibiting the desired properties. 
When the driving frequency is large enough compared to the characteristic energy scales of the system, 
the effective Hamiltonian evaluated with the high frequency expansion describes the system for exponentially long times \cite{Mori2016,Kuwahara2016}. 
However, for nonintegrable systems and not too large driving frequency, the validity of these assumptions is not a priori clear. 
Such systems are expected to heat up in the presence of periodic driving, and it is an interesting question whether, and for how long, a quasi-steady 
``Floquet prethermalized state" (FPS)  
different from the trivial infinite temperature state can be established.

\begin{figure*}[ht]
\centering
\includegraphics[width=0.95\textwidth]{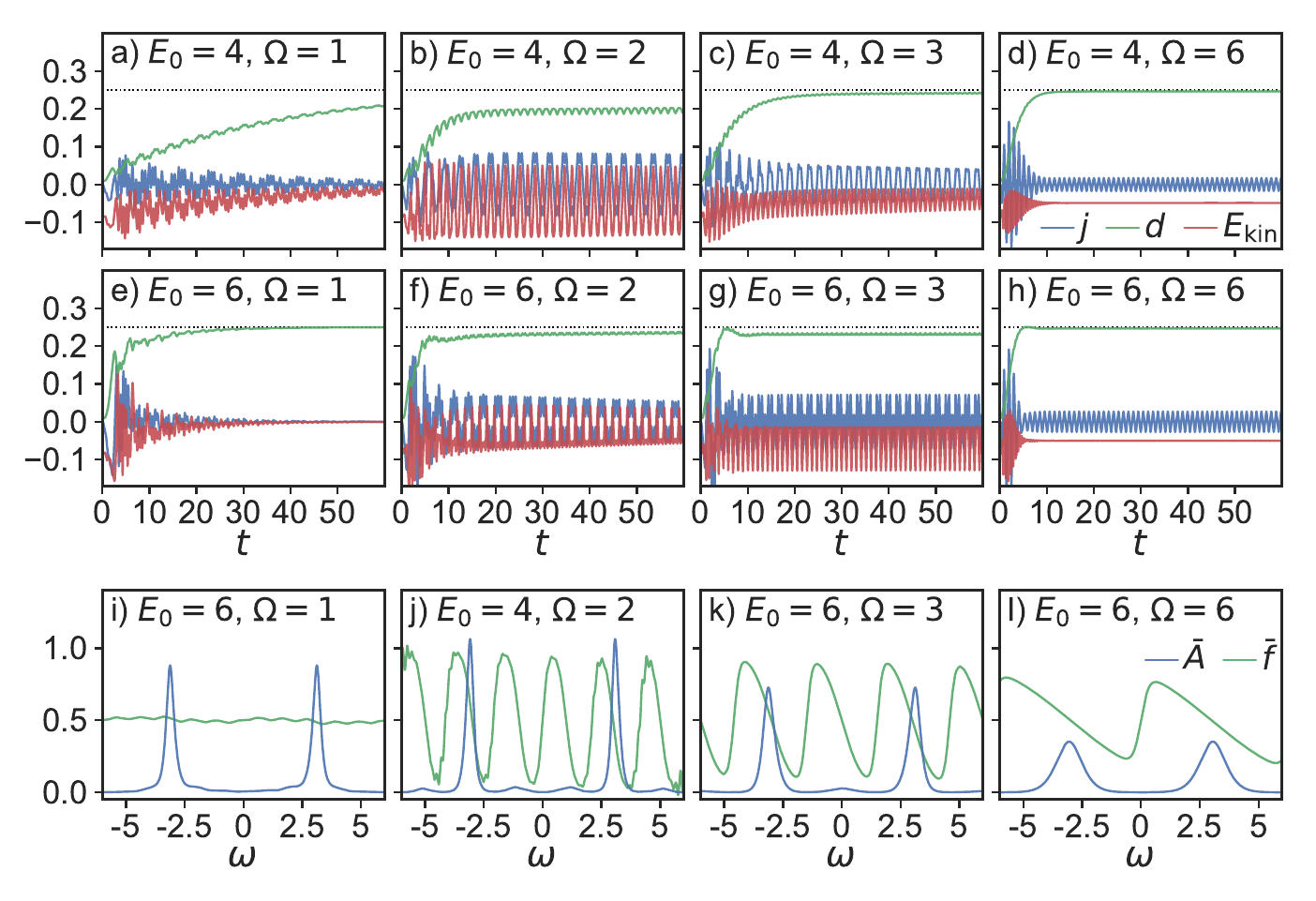}
\caption{Half-filled Hubbard model with $U=6$ and initial inverse temperature $\beta=2$. Panels a)-h): Time evolution of the current $j$, double occupation $d$, and kinetic energy $E_\text{kin}$ after the switch-on of an electric field with indicated amplitude $E_0$ and frequency $\Omega$.
Panels i)-l): Time averaged nonequilibrium spectral function $\bar A(\omega)$ and nonequilibrium distribution function $\bar f(\omega)$ 
for the indicated values of $E_0$ and $\Omega$. 
}
\label{fig1}
\end{figure*}

Several recent theoretical works have demonstrated the existence of long-lived 
FPSs  
in interacting models  \cite{Mori2016,Dalessio2013,Citro2015,Bukov2015b,Canovi2016,Weidinger2017}, or questioned the general belief that heating to infinite temperature occurs in such systems. 
In this work we consider the Mott insulating single-band Hubbard model in time-periodic electric fields.
We will show that 
FPSs exist under various driving conditions, even when the driving frequency is smaller than the Hubbard interaction and comparable to the bandwidth. 
Focusing specifically on the case of resonant driving, where the local interaction is a multiple of the driving frequency and a violent heating 
might be naively expected, we demonstrate that the system can be trapped in long-lived 
states 
with a suppressed number of double occupations and/or a nonzero current and kinetic energy.  
In the driving regime 
where the leading order term in $H_\text{eff}$ describes free fermions, 
we can interpret the FPS as a  
state resulting from 
a quench to a weakly interacting effective Hubbard model.  

{\it Model.} The Hamiltonian of the driven Hubbard model is 
$H(t)=-v\sum_{\langle i,j\rangle\sigma} c^\dagger_{i\sigma}c_{j\sigma}+U\sum_j (n_{j\uparrow}-\tfrac{1}{2})(n_{j\downarrow}-\tfrac{1}{2})-qE(t)\sum_{i\sigma}(\hat e \cdot \vec{r}_i)n_{i\sigma}$ 
with $c^\dagger_{i\sigma}$ the creation operator for an electron of spin $\sigma$ at site $i$, $n=c^\dagger c$ the number operator, $v$ the nearest-neighbor hopping, $U$ the on-site interaction, $q$ the electron charge (which is set to 1), and $E \hat e$ the electric field with polarization $\hat e$. 
To solve the model we use the nonequilibrium dynamical mean field theory (DMFT) \cite{Aoki2014} in combination with a strong coupling perturbative impurity solver (noncrossing approximation, NCA) \cite{Keiter1971,Eckstein2010}. We consider an infinite-dimensional hybercubic lattice 
with a Gaussian density of states $\rho(\epsilon)=\exp(-\epsilon^2/W^2)/\sqrt{\pi} W$  
and apply the electric field along the body diagonal, $E(t) \hat e=(E(t), E(t), \ldots)$. In a gauge with pure vector potential $A(t)$, the electric field $E(t)=-\partial_t A(t)$ enters the calculation as a time-dependent shift of the noninteracting dispersion, $\epsilon_k\rightarrow \epsilon_{k-qA(t)}$. The implementation and NCA treatment of this problem has been discussed in Refs.~\cite{Eckstein2011,Aoki2014}. We start at $t=0$ in the equilibrium state at inverse temperature $\beta$ and 
switch on the electric field as $E(t)=E_0 \sin(\Omega t)$,   
where $E_0$ is the field amplitude and $\Omega$ the driving frequency. 
Energy is measured in units of $W=1$  and time in units of $W^{-1}$. 

{\it Results.} Figure~\ref{fig1} shows the time dependence of the current $j$, the double occupation $d$ and the kinetic energy $E_\text{kin}$ in a model with $U=6$ for indicated values of the field amplitude and driving frequency. (For the DMFT measurement of these quantities, see Ref.~\cite{Aoki2014}.) 
All these results correspond to resonant driving ($U=n\Omega$, with $n$ integer). In the non-resonant case, the time evolution is slow, and we cannot reach the timescales needed to observe a saturation in a prethermalized state, or a heating to infinite temperature. 
 In the following, we will thus focus on resonantly driven systems, where the doublon-holon production is strong and a (quasi-)steady 
 FPS  
is rapidly reached. 

Panels a) and e) show that for small driving frequency and large field amplitude, the system quickly approaches the infinite temperature state characterized by $j=0$, $d=0.25$ and $E_\text{kin}=0$. 
Especially for the stronger field ($E_0=6$) the main mechanism for doublon-holon production in this low-frequency driving regime is field-induced tunneling, which creates an almost flat (infinite temperature) energy distribution of the photo carriers \cite{Oka2012,Eckstein2013}. 
On the other hand, for $\Omega=2$ and $3$, the system can be trapped in a long-lived quasi-steady state with a suppressed double occupation and a strongly oscillating current and kinetic energy (panels b), f), c) and g)). 
Note that the drift of the double occupation and kinetic energy, and hence the heating rate in the quasi-steady state, depends in a nontrivial way on the field amplitude and driving frequency. 
There are also examples of FPSs 
with 
$d\approx 0.25$, but $E_\text{kin}\approx \text{const} < 0$ (panels d) and h)).   
In general, the time scale on which the steady state (either infinite temperature state or FPS) is reached becomes longer as we decrease the field strength. 

The effective temperature of the doublons and holons in the trapped state can be 
estimated by calculating  
the time-dependent spectral functions from the
lesser and retarded local Green's function $G$ as $A^<(\omega,t)=\frac{1}{2\pi}\text{Im}\int_t^{t_\text{max}} dt' e^{i\omega(t'-t)}G^<(t',t)$ and $A(\omega,t)=-\frac{1}{\pi}\text{Im}\int_t^{t_\text{max}} dt' e^{i\omega(t'-t)}G^\text{ret}(t',t)$. 
Averaging these spectral functions over one period yields $\bar A^<(\omega)$ and $\bar A(\omega)$,  
which can be used to define the 
``nonequilibrium distribution function" $\bar{f}(\omega)=\bar A^<(\omega)/\bar A(\omega)$. 
Panels i)-l) of Fig.~\ref{fig1} show $\bar A(\omega)$ and $\bar{f}(\omega)$  
obtained by averaging the Fourier transforms starting from $t=25$. 
The almost flat distribution function in panel i) confirms the expectation from the quasi-static picture. 
Fitting the slopes of the distribution functions in panels j)-l) with a Fermi function $1/(1+\exp(\beta_\text{eff}(\omega-\mu_\text{eff})))$ 
in the energy range of the Hubbard bands
yields $\beta_\text{eff}=5.53$, $\mu_\text{eff}=\pm 3.00$ for $E_0=4,\Omega=2$, $\beta_\text{eff}=2.61$, $\mu_\text{eff}=\pm 3.00$ for $E_0=6,\Omega=3$, and $\beta_\text{eff}=0.495$, $\mu_\text{eff}=\pm 3.00$ for $E_0=4,\Omega=6$. 
Hence, the effective temperature of the long-lived trapped state can be lower than the temperature of the initial equilibrium state ($\beta=2$). For driving frequencies $\Omega$ somewhat above the resonant value $U/n$ we furthermore find negative effective temperatures, i.e. inverted populations of doublons and holons in the Hubbard bands.

We also note that $\bar f(\omega)$ satisfies $\bar f(\omega+n\Omega)=\bar f(\omega)$. 
This property can be explained by analyzing the Green's functions in terms of the kick operator and the effective Hamiltonian assuming that the system has reached a thermal state of the effective Hamiltonian at $\beta_{\rm eff}$ and that the characteristic energy scale of the effective Hamiltonian is smaller than $\Omega$. 
These assumptions imply 
that $\bar{f}(\omega)$ around $\omega=n\Omega$ behaves like the Fermi distribution function $f(\omega-n\Omega)$ at $\beta_{\rm eff}$, see Supplemental Material for details.  

While the response of the Mott insulating Hubbard model to the electric field driving is complex and depends sensitively on the parameters of the driving field, our DMFT results clearly demonstrate the existence of long-lived 
FPSs, even for resonant driving, moderate frequencies ($\Omega\gtrsim W$) and for large field amplitudes. 
An interesting question is how well the FPSs and the relaxation into these states are described by the  
time-independent effective Hamiltonian. 

\begin{figure}[t]
\includegraphics[width=\columnwidth]{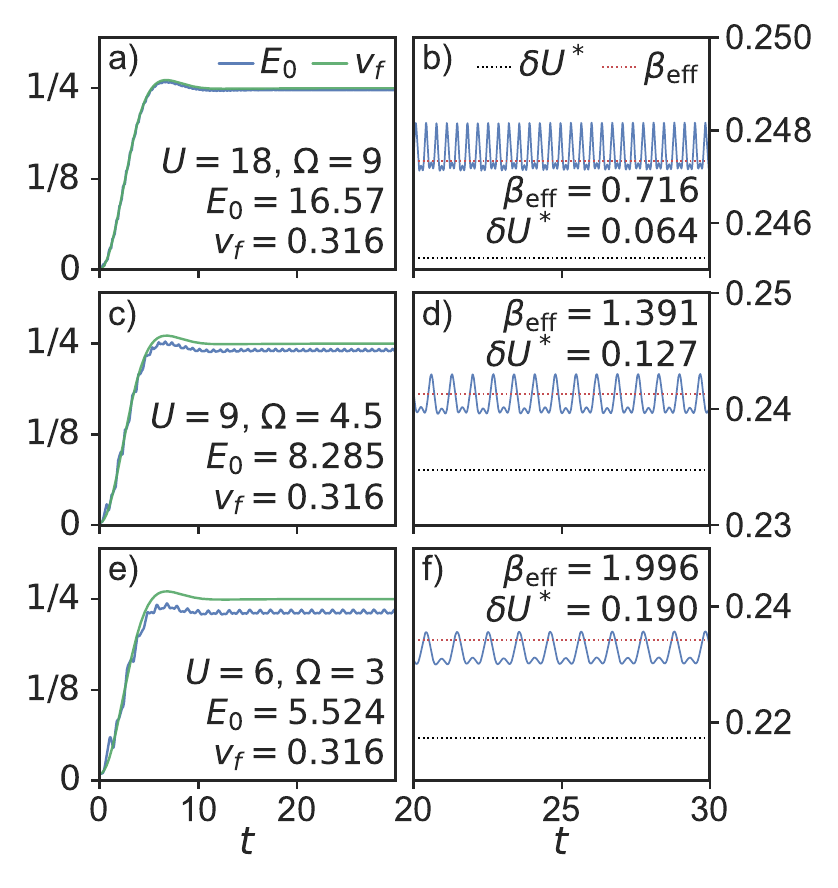}
\caption{Resonant driving with $n=2$. Panels a), c) and e) compare the time-evolution of the driven Hubbard model (blue) to the quench dynamics for a quench to the leading order (noninteracting) static model $H_\text{eff}^{(0)}$. Here, the driving amplitude is chosen such that $\mathcal{J}_0(E_0/\Omega)=\mathcal{J}_2(E_0/\Omega)$ corresponds to the first crossing of the Bessel functions, and the final hopping is $v_f=v\mathcal{J}_0(E_0/\Omega)$. Panels b), d), f): Zoom of the plateau region and comparison with the $U$-quench to $|\delta U^*|$ and simultaneous hopping quench to $v_f$ (black line), as well as the prediction from the equilibrium interacting model with $U=|\delta U^*|$  and inverse temperature identical to $\beta_\text{eff}$ of the driven state (red line).
}
\label{fig2}
\end{figure}

To shed some light on this issue, we consider the case of resonant driving $U=n\Omega$ with even $n$, where the effective Hamiltonian becomes simple. As shown by Bukov, Kolodrubetz and Polkovnikov \cite{Bukov2016}, the leading-order effective Hamiltonian $H^{(0)}_\text{eff}$ has two terms, describing doublon/holon hopping, and doublon-holon production/recombination. 
For even $n$ and suitably chosen driving amplitude, $H^{(0)}_\text{eff}$ corresponds to free fermions. In the specific case of a hypercubic lattice with hopping $v$, 
the leading term in the high-frequency expansion becomes 
$H^{(0)}_\text{eff}=-v\sum_{\langle i,j\rangle\sigma}[\mathcal{J}_0(\xi)g_{ij\sigma}+(\mathcal{J}_{n}(\xi)(-\eta_{ij})^n h^\dagger_{ij\sigma} + h.c.)]$, with $\xi=|Z_{ij}|$, $\eta_{ij}=\text{sign}(Z_{ij})$ and $Z_{ij}=\frac{-qE_0}{\Omega}\hat e\cdot \vec r_{i-j}$ ($q=1$). 
Here, $\mathcal{J}_n$ denotes the $n$th order Bessel function, $g_{ij\sigma}=(1-n_{i\bar\sigma})c^\dagger_{i\sigma}c_{j\sigma}(1-n_{j\bar\sigma})+n_{i\bar\sigma}c^\dagger_{i\sigma}c_{j\sigma}n_{j\bar\sigma}$ the operator describing the hopping of doublons and holons, and $h^\dagger_{ij\sigma}=n_{i\bar\sigma}c^\dagger_{i\sigma}c_{j\sigma}(1-n_{j\bar\sigma})$ the operator for doublon-holon production.\footnote{To be precise, this effective Hamiltonian is for $E(t)=E_0\cos(\Omega t)$. In our case of $E(t)=E_0\sin(\Omega t)$, the leading order effective Hamiltonian from the van Vleck high-frequency expansion is identical to the free Hamiltonian after a further unitary transformation, which can be absorbed into a redefinition of the kick operator.} If $\frac{E_0}{\Omega}$ is chosen such that the two amplitudes are equal, i.e. $\mathcal{J}_0(\xi)=\mathcal{J}_n(\xi)$, and $n$ is even so that $(-\eta_{ij})^n=1$, then the driven system (in the high-frequency limit) is expected to behave like a noninteracting model with hopping amplitude $\mathcal{J}_0(\xi)v$. 

We demonstrate this behavior in Fig.~\ref{fig2}, where we compare the time evolution of the double occupation of the driven system with interaction $U$ to the double occupation in a Hubbard model after a quench 
from $U_\text{initial}=U$ to $U_\text{final}=0$ and a simultaneous reduction in the hopping amplitude.  
We choose $\Omega=U/2$ and $E_0/\Omega=1.841$, 
so that  $\mathcal{J}_0(E_0/\Omega)=\mathcal{J}_2(E_0/\Omega)=0.316$, and we 
quench the hopping amplitude from $v_\text{initial}=v$ to $v_\text{final}=0.316 v$. 
For $U=18$ (panel a)) we are in the high frequency driving regime, where the leading order effective Hamiltonian of Bukov {\it et al.} should be valid. Indeed, the time dependence of the double occupation shows the behavior expected for a quench from the undriven $H$ to $H_\text{eff}^{(0)}$ and the double occupation increases to a value close to $d=0.25$. 
As the interaction (and hence the driving frequency) is reduced (panels c) and e)), larger deviations between the driven system and the effective static description appear. While the quench  
to the noninteracting $H_\text{eff}^{(0)}$ inevitably leads to a saturation of the double occupation at $d=0.25$, the dynamics of the driven system shows a trapping of $d$ at a value below 0.25. 

\begin{figure}[ht]
\includegraphics[width=\columnwidth]{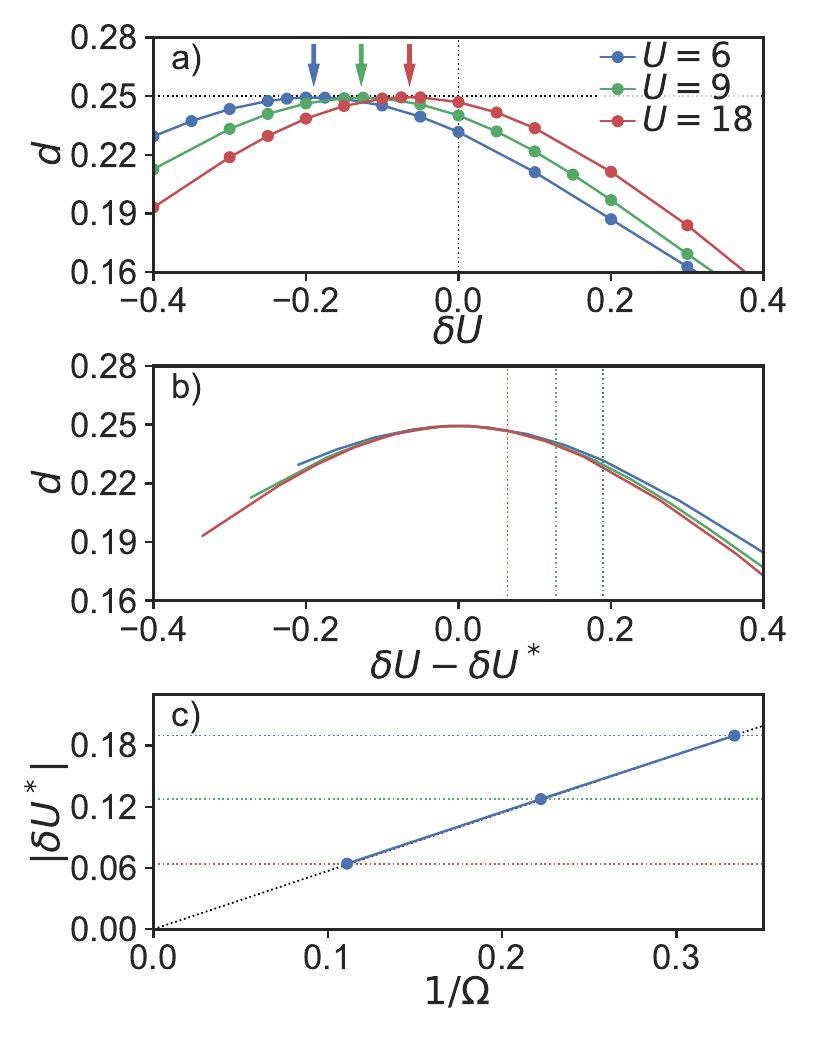}
\caption{
Resonantly driven Hubbard model with interaction $U+\delta U$, driving frequency $\Omega=U/2$ and amplitude $E_0/\Omega=1.841$ (initial inverse temperature $\beta=2$). Panel a): double occupation in the FPS for different values of $U$. The arrows indicate the value $\delta U^*$, where the double occupation reaches a maximum close to 0.25.  
Panel b): double occupation as a function of $\delta U-\delta U^*$ with dashed lines at $|\delta U^*|$. 
Panel c): scaling of $|\delta U^*|$ with inverse driving frequency.
}
\label{fig3}
\end{figure}

\begin{figure}[ht]
\includegraphics[width=\columnwidth]{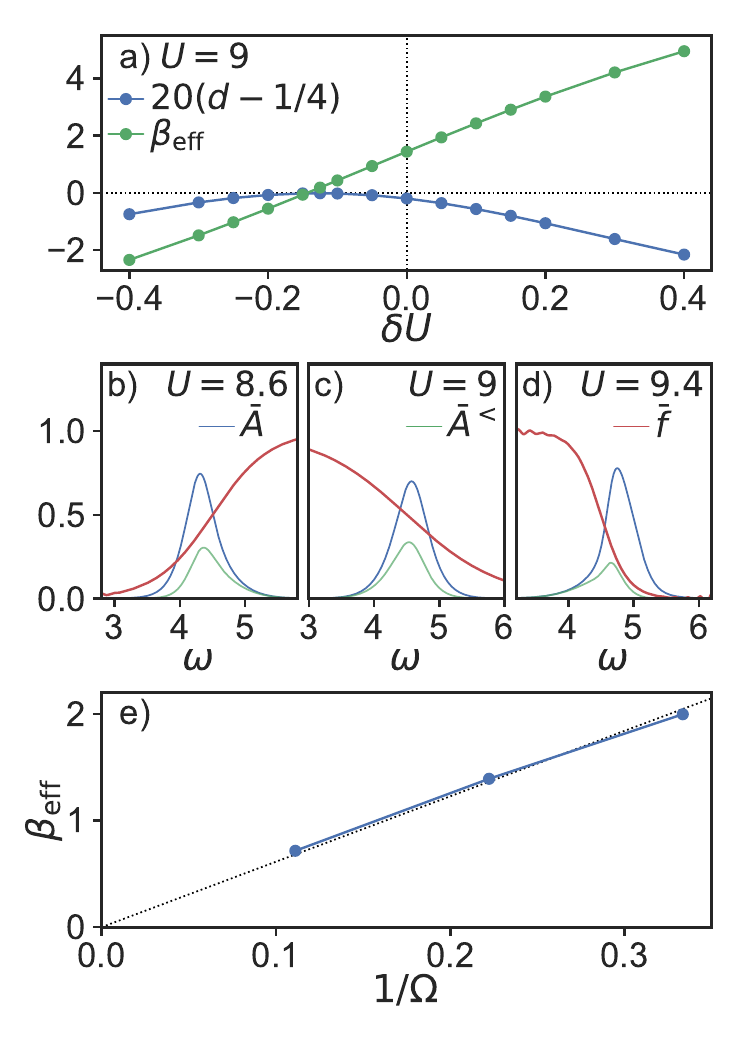}
\caption{
Resonantly driven Hubbard model with interaction $U+\delta U$, driving frequency $\Omega=U/2$ and amplitude $E_0/\Omega=1.841$ ($U=9$, initial inverse temperature $\beta=2$). Panel a): effective inverse temperature $\beta_\text{eff}$ of the doublons and holons. 
Panels b-d): Time-averaged spectral functions $\bar A$, $\bar A^<$ and energy distribution functions $\bar f(\omega)$ of the driven system for $\delta U=+0.4$, $0$, $-0.4$. Panel e): $\beta_\text{eff}$ for $\delta U=0$ as a function of inverse driving frequency ($U=6$, $9$, $18$). 
}
\label{fig4}
\end{figure}

The observed deviations from the effective model description must be due to the higher order corrections in $H_\text{eff}$. 
We have calculated $H_\text{eff}^{(1)}$ (see Supplemental Material), and it contains a large number of terms involving up to three different sites.
One can see that $H_\text{eff}^{(1)}$ induces 
correlations and modifications of the bandwidth, which represent the O($\frac{1}{\Omega}$) corrections to the leading order noninteracting Hamiltonian $H_\text{eff}^{(0)}$. 
For example, some of these terms describe the hopping of a pair of electrons from (to) the same site. This acts like a local interaction whose strength is determined by some average kinetic energy squared times a prefactor $\sim \frac{1}{\Omega}$.  
Another term describes a three body interaction $n_{i\uparrow}n_{i\downarrow}\bar{n}_{j}$, which may also act as a local interaction whose strength is determined by the average of the occupancy on neighbouring sites. 
In addition there are correlated hopping terms which cannot be reduced to an effective Hubbard interaction, but which may change the effective bandwidth. 
It is thus an interesting question to what extent the effect of the additional terms in $H_\text{eff}$ can be captured by a simple Hubbard Hamiltonian with modified interaction and bandwidth. In the following, we will demonstrate that to a large extent, $H_\text{eff}$ acts as a local Hubbard interaction, and we will use this insight to interpret the FPSs with $d<0.25$ shown in Fig.~\ref{fig2}.

First of all, we note that a significant change of the effective bandwidth would manifest itself in a change of the timescale on which the double occupation grows after the electric field quench. However, Fig.~\ref{fig2} shows that the quench $v\rightarrow 0.316v$ correctly reproduces this growth rate not only in the high-frequency regime ($U=18$), but also for $U=9$ and $6$. This implies that the band widening effect of $H_\text{eff}^{(1)}$ is not significant. 

Instead, our numerical analysis shows that the effects of $H_\text{eff}$ are, to a large extent, mimicked by a local Hubbard interaction $U_\text{eff}$. 
To determine $U_\text{eff}$, 
we drive the system with $\Omega=U/2$, choose $E_0/\Omega=1.841$ corresponding to the noninteracting condition for $H_\text{eff}^{(0)}$, and vary the interaction of the driven system as $U'=U+\delta U$. In the high-frequency driving regime, this model should behave like a static Hubbard model with interaction $\delta U$ and a rescaled hopping parameter. At finite driving frequency, $H_\text{eff}^{(1)}$ produces additional interaction effects, so that the effective static model is $H_\text{eff}=H_\text{eff}^{(0)}+\sum_i \delta U (n_{j\uparrow}-\tfrac{1}{2})(n_{j\downarrow}-\tfrac{1}{2})+H_\text{eff}^{(1)}+O(\frac{1}{\Omega^2})$. The measured time-averaged double occupation in the trapped state is plotted as a function of $\delta U$ in the top panel of Fig. 3. The double occupation reaches a maximum very close to the noninteracting value $d=0.25$ at some negative value $\delta U^*$ of $\delta U$. 
There is apparently a cancellation between the interaction coming from $H_\text{eff}^{(1)}$ and the local interaction $\delta U$ at this particular point. Conversely, the resonantly driven system with $\delta U=0$ has an effective Hubbard interaction of strength $U_{\rm eff}=|\delta U^*|$, which originates from $H_\text{eff}^{(1)}$.
Indeed $|\delta U^*|$ scales with $1/\Omega$ (see panel c)), which is consistent with the interpretation of an effective interaction coming from the next leading order.

In panel b) of Fig.~\ref{fig3}, we plot the double occupation in the FPS as a function of $\delta U-\delta U^*$. 
If the effective Hubbard interaction $|\delta U^*|$ would perfectly capture the effect of higher-order terms in $H_\text{eff}$, 
and the absorbed energy were independent of $\Omega$ and the kick operator, 
we would expect a collapse of the curves for different $U$. 
The shifted data show a rather good agreement for $U=18$, $9$ and $6$, but small deviations remain.  
These deviations indicate that some of the correlations induced by $H_\text{eff}^{(1)}$ cannot be captured by an effective Hubbard interaction, and they provide a rough estimate of these beyond-Hubbard effects on $d$.

An interesting question is why the double occupation shows a parabolic maximum near $d=0.25$ as the interaction is varied near the resonance condition (Fig. 3a)). 
In fact, the expression for $H_\text{eff}$ suggests that below the effective noninteracting point $\delta U^*$, the driven system should behave like an attractive Hubbard model, 
which usually yields $d>0.25$.
The observed suppression for $\delta U<\delta U^*$ occurs because in this regime, the driven system exhibits an {\it inverted population}. 
In panel a) of Fig.~\ref{fig4} we plot the effective inverse temperature of the model with $U=9$ as a function of $\delta U$. 
 This figure shows that the effectively noninteracting driven system has an infinite temperature distribution, while the effectively attractive system has a negative effective temperature.
 The nonequilibrium distribution functions $\bar f(\omega)$ for $\delta U=+0.4$, $0$, $-0.4$ are illustrated in panel b). 
  As discussed in Ref.~\cite{Tsuji2011}, the Hubbard model with interaction 
$\delta U-\delta U^*$ and negative effective temperature can be mapped onto a Hubbard model with interaction $-(\delta U-\delta U^*)$ and positive temperature. 
This explains why the doublon occupation does not exceed 0.25 even when the effective model shows an attractive interaction. 
On the other hand, the resonantly driven model ($\delta U=0$) with effective interaction $U_\text{eff}=|\delta U^*|$ coming from $H_\text{eff}^{(1)}$ has a positive temperature distribution 
and an effective inverse temperature comparable to the initial equilibrium state ($\beta=2$). 
For driving frequencies below the resonance ($\delta U>0$), 
the driven system can be effectively colder than the initial state. 

We also note that the parabolic dependence of $d$ on $\delta U$ and the $|\delta U^*|\sim \frac{1}{\Omega}$ scaling imply that the deviation of the double occupation in the FPS from 0.25 is proportional to $\frac{1}{\Omega^2}$. 
This at first sight surprising scaling is the result of an effective interaction proportional to $1/\Omega$ and the fact that the system approaches an infinite temperature state in the high-frequency limit as $\beta_{\rm eff}\propto1/\Omega$, see Fig.~\ref{fig4}e).

It is interesting to check how well the effective Hubbard model   
explains the observed values of the double occupation in the FPSs of the resonantly driven system (Fig.~\ref{fig2}). An interaction quench 
to $|\delta U^*|$ 
does not reproduce the plateau value very well (black dashed line). This is because the absorbed energy, and hence the effective temperature of the trapped or quenched state, depends sensitively on the details of the transient evolution, 
i.e. the kick operator. 
It is thus more meaningful to extract the effective inverse temperature $\beta_\text{eff}$ of the driven system from a Fermi function fit of $\bar f(\omega)$ and to compare the double occupation of a Hubbard model with interaction $|\delta U^*|$ and inverse temperature $\beta_\text{eff}$ to the double occupation in the FPS. The corresponding results are indicated by the red dashed lines in the right hand panels of Fig.~\ref{fig2},
and they are in rather good agreement with the time-averaged double occupation. 
 The remaining deviation to the FPS is comparable to the deviations evident in Fig.~\ref{fig3}b), and may be attributed to ``beyond-Hubbard" interaction effects. 

We finally comment on the question whether the FPS observed here is a thermal or prethermal state in terms of $H_\text{eff}$. 
Due to the vicinity to the integrable noninteracting limit, the FPS might be expected to be a long-lived {\it prethermalized state} \cite{Moeckel2008}, where the properties of nonlocal observables are different from those of the thermalized system described by $H_\text{eff}$.
Though the direct simulation of $H_{\rm eff}$ and the analysis of nonlocal observables are beyond the scope of this study, we confirmed that the quench to the effective Hubbard model with $U_\text{eff}=|\delta U|$ shows a fast thermalization of local observables such as the double occupation, kinetic energy, and distribution function $f(\omega)$. 

{\it Summary.} 
In this study, we have analyzed the properties of the resonantly driven Mott insulating Hubbard model. Contrary to naive expectations, and despite an efficient doublon-holon production in the resonant regime, this nonintegrable system can be trapped in long-lived 
Floquet prethermal 
states characterized by a suppressed double occupation and a nonzero current. While such trapping phenomena are found under various driving conditions, we have focused on the case $U=n\Omega$, with $n$ even, where the leading order effective Hamiltonian reduces to a noninteracting fermion model. In this driving regime, the long-lived trapped states can be understood as 
states resulting from a quench to a weakly interacting effective Hamiltonian. While these interactions originate from higher order terms in the high-frequency expansion, i.e., multi-site correlated hopping terms, their effect on local observables 
can to a large extent be captured by an effective Hubbard repulsion $U_\text{eff}=|\delta U^*|$. We have demonstrated a numerical procedure for evaluating $|\delta U^*|$ and showed that it scales with $1/\Omega$, as expected for an interaction resulting from the next-leading order.   
At driving frequencies comparable to the bandwidth, there are however non-negligible beyond-Hubbard interaction effects. 

We have also demonstrated 
that driving below the resonance can lead to a cooling of the doublons and holons, and 
that for driving above the resonance, the  driven Mott insulators can exhibit inverted doublon and holon populations. In the latter case, a sign change in the interaction terms of the effective Hamiltonian is required to correctly describe the properties of the Floquet prethermalized states by means of the effective static model 
with a positive temperature.     

{\it Acknowledgments} The calculations have been performed on the Beo04 cluster at the University of Fribourg. This work was supported by the European Research Council through ERC Starting Grant 278023 and Consolidator Grant 724103, and by the Swiss National Science Foundation through NCCR Marvel.

\bibliographystyle{eplbib}

%%%%%%%%%%%%%%%%%%%%%%%%%

\newpage

\onecolumn

\noindent
{\bf Supplemental material: Floquet prethermalization in the resonantly driven Hubbard model}\\

\noindent
{\it A.1. General idea of the Schrieffer-Wolff transformation for periodically driven systems}\\

In our study, we follow Ref.~\cite{Bukov2016} to evaluate the effective Hamiltonian for the AC driven Hubbard model
in the resonant or nearly resonant regime. 
In a rotating frame, the Hamiltonian becomes time dependent with the characteristic frequencies of 
the interaction ($U$) and the excitation frequency ($\Omega$). 
We can then define a common energy scale $\Omega_0$ of these two and perform a high-frequency expansion if $\Omega_0$ is much higher than the kinetic energy.  
Without the driving this procedure leads to the Schrieffer-Wolff transformation for the $t$-$J$ model or the Heisenberg model.

To be more precise, our Hamiltonian is the Hubbard model under a periodic excitation of the form 
\begin{align}
H(t)=&-v \sum_{\langle ij\rangle,\sigma} c^\dagger_{i\sigma}c_{j\sigma}+(U+\delta U)\sum_j n_{j\uparrow}n_{j\downarrow}+\sum_{j\sigma}f_{j\sigma}(t)n_{j\sigma}.
\end{align}
The rotating frame with respect to $V(t)=e^{-i[Ut\sum_j n_{j\uparrow} n_{j\downarrow}+\sum_{j\sigma}F_{j\sigma}(t)n_{j\sigma}}]$, where 
$F_{j\sigma}(t)=\int^t_0 f_{j\sigma}(t')dt'$, is chosen to obtain
\begin{align}
&H^{\rm rot}(t)=V^\dagger(t) H(t)V(t)=-v\sum_{\langle ij\rangle \sigma}[e^{i\delta F_{ij\sigma}(t)}g_{ij\sigma}+(e^{i\delta F_{ij\sigma}(t)+iUt}h^\dagger_{ij\sigma}+h.c.)]+\delta U \sum_j n_{j\uparrow}n_{j\downarrow},
\end{align}
where
\begin{align}
&g_{ij\sigma}=\bar{n}_{i\bar{\sigma}}c^\dagger_{i\sigma}c_{j\sigma}\bar{n}_{j\bar{\sigma}}
+n_{i\bar{\sigma}}c^\dagger_{i\sigma}c_{j\sigma}n_{j\bar{\sigma}},\quad h_{ij\sigma}^\dagger=n_{i\bar{\sigma}}c^\dagger_{i\sigma}c_{j\sigma}\bar{n}_{j\bar{\sigma}}.
\end{align}
Here $\bar{n}_{i\bar{\sigma}}=1-n_{i\bar{\sigma}}$, 
$g_{ij\sigma}$ represents the hopping of doublons and holons and $h_{ij\sigma}^\dagger$ the creation of doublon-holon pairs.

The second step is to regard this Hamiltonian as a time-periodic Hamiltonian and to perform a high-frequency expansion.
When the Hamiltonian is time-periodic, 
from the Floquet theorem, the full time evolution $U(t_2,t_1)=\mathcal{T}\exp(-i\int^{t_2}_{t_1}H(t)dt)$ can be decomposed as 
$U(t_2,t_1)=\exp[-iK_{\rm eff}(t_2)]\exp[-iH_{\rm eff}(t_2-t_1)]\exp[iK_{\rm eff}(t_1)]$.
Here $H_{\rm eff}$ is the time-independent effective Hamiltonian and $K_{\rm eff}(t)$ is the kick operator, which is time periodic with period $T=2\pi/\Omega$.
Usually, these operators can be approximately evaluated with high-frequency expansions.
We note that the expansion is not unique and several methods exist \cite{Mikami2016}.
One possible procedure is the van Vleck high-frequency expansion \cite{Bukov2016}. 

To apply the high-frequency expansion, 
we introduce a common frequency $\Omega_0$ so that $\Omega=k_0\Omega_0$ and $U=m_0\Omega_0$
and write
\begin{align}
e^{i\delta F_{ij\sigma}(t)}&=\sum_l A^{(l)}_{ij\sigma}e^{il\Omega_0 t},\nonumber\\
e^{i\delta F_{ij\sigma}(t)+Ut}&=\sum_{l}A^{(l)}_{ij\sigma}e^{i(l+m_0)\Omega_0 t}=\sum_{l} B^{(l)}_{ij\sigma}e^{il\Omega_0 t}.
\end{align}
From the van Vleck high-frequency expansion, the leading terms of $H_{\rm eff}$ become 
\begin{align}
H^{(0)}_{\rm eff}=&-v\sum_{\langle ij\rangle \sigma}[A^{(0)}_{ij}g_{ij\sigma}+(B^{(0)}_{ij\sigma}h_{ij\sigma}^\dagger+B^{(0)*}_{ij\sigma}h_{ij\sigma})]+\delta U \sum_j n_{j\uparrow}n_{j\downarrow},\\
H^{(1)}_{\rm eff}=&\frac{1}{\Omega_0}\sum_{l=1}^\infty \frac{1}{l}[H^{\rm rot}_l,H^{\rm rot}_{-l}],
\end{align}
where  $H^{\rm rot}_l=-v\sum_{\langle ij\rangle \sigma}[A^{(l)}_{ij\sigma}g_{ij\sigma}
+(B^{(l)}_{ij\sigma}h_{ij\sigma}^\dagger+B^{(-l)*}_{ij\sigma}h_{ij\sigma})]$.
We also note that 
in these expressions, 
we can reduce some terms 
under the assumption 
that the number of doublons and holons is small. \\

\noindent
{\it A.2. Application to the resonantly driven Mott insulator}\\

Now we apply this theory to our case of the infinite-dimension hyper-cubic lattice under AC driving near the resonance.
This is the case when $k_0=1$. Then one can resonantly excite the system using multi-photon processes. 
Therefore one cannot neglect the terms involving doublons or holons.

The driving term explicitly reads
\begin{align}
\sum_{j\sigma}f_{j\sigma}(t)n_{j\sigma}=-qE(t)\sum_{i,\sigma} ({\bf e}_0\cdot {\bf r}_i) n_{i,\sigma},
\end{align}
where ${\bf e}_0=(1,1,1,\cdots,1,1)$.
 For $E(t)=E_0\cos(\Omega t)$, $\delta F_{i,j}(t)=-\frac{qE_0}{\Omega}\sin (\Omega t) {\bf e}_0\cdot {\bf r}_{i-j}\equiv Z_{ij}\sin (\Omega t)$.
From this one can see that
\begin{align}
A^{(l)}_{ij}=\mathcal{J}_{-l}(-Z_{ij})=\mathcal{J}_{l}(Z_{ij})=\mathcal{J}_{l}(\zeta)\eta_{ij}^l.\label{eq:coeffs}
\end{align}
Here $\zeta=|Z_{ij}|$ and $\eta_{ij}={\rm sign}(Z_{ij})$ and $\mathcal{J}_l$ denotes the $l$-th order Bessel function.
For $E(t)=E_0\sin(\Omega t)$, the effective Hamiltonian is identical to the case of $E(t)=E_0\cos(\Omega t)$ after the unitary transformation $\Theta=\exp[i\sum_{j\sigma}\frac{qE_0}{\Omega}{\bf e}_0\cdot {\bf r}_j n_{j\sigma}-iU\frac{T}{4} \sum_j n_{j\uparrow}n_{j\downarrow}]$.

The leading order becomes 
\begin{align}
H^{(0)}_{\rm eff}=&-v\sum_{\langle ij\rangle \sigma}[\mathcal{J}_{0}(\zeta)g_{ij\sigma}+(\mathcal{J}_{-m_0}(\zeta)\eta_{ij}^{m_0}h_{ij\sigma}^\dagger+h.c.)]\nonumber\\
&+\delta U \sum_j n_{j\uparrow}n_{j\downarrow}.
\end{align}
Hence, the leading term describes correlated hoppings, and 
one can see that at some special points the system becomes effectively free.

We now calculate the next order correction $H_\text{eff}^{(1)}$. 
One can identify three different groups of terms. 
\begin{itemize}
\item Group 1\\
\begin{align}
H^{(1)}_{\rm eff,1}=\frac{1}{2}\sum_{l\neq 0}\frac{v^2}{l\Omega}
 \Bigg[\sum_{\langle ij\rangle \sigma}A^{(l)}_{ij\sigma}g_{ij\sigma},\sum_{\langle i'j'\rangle \sigma'}A^{(-l)}_{i'j'\sigma'}g_{i'j'\sigma'}\Bigg].
\end{align}
These terms vanish in our case of AC field driving because $\sum_{\langle ij\rangle \sigma}A^{(l)}_{ij}g_{ij\sigma}\propto\sum_{\langle i'j'\rangle \sigma'}A^{(-l)}_{i'j'}g_{i'j'\sigma'}$.

\item Group 2\\
\begin{align}
H^{(1)}_{\rm eff,2}&=\sum_{l\neq 0}\frac{v^2}{l\Omega}
 \Bigg[\sum_{\langle ij\rangle \sigma}A^{(l)}_{ij\sigma}g_{ij\sigma},\sum_{\langle i'j'\rangle \sigma'}B^{(-l)}_{i'j'}h^\dagger_{i'j'\sigma'}+B^{(l)*}_{i'j'\sigma'}h_{i'j'\sigma'}\Bigg].
\end{align}

\item Group 3\\
\begin{align}
H^{(1)}_{\rm eff,3}&=\frac{1}{2}\sum_{l\neq 0}\frac{v^2}{l\Omega}
 \Bigg[\sum_{\langle ij\rangle \sigma}B^{(l)}_{ij}h^\dagger_{ij\sigma}+B^{(-l)*}_{ij\sigma}h_{ij\sigma},\sum_{\langle i'j'\rangle \sigma'}B^{(-l)}_{i'j'}h^\dagger_{i'j'\sigma'}+B^{(l)*}_{i'j'\sigma'}h_{i'j'\sigma'}\Bigg]\nonumber\\
 &=\sum_{l\neq 0}\frac{v^2}{l\Omega}
\Bigg[\sum_{\langle ij\rangle \sigma}B^{(l)}_{ij}h^\dagger_{ij\sigma},\sum_{\langle i'j'\rangle \sigma'}B^{(l)*}_{i'j'\sigma'}h_{i'j'\sigma'}\Bigg].
 \end{align}
Here we can directly show that $[h^\dagger_{ij\sigma},h^\dagger_{i'j'\sigma'}]=0$.
\end{itemize}

Let us write down the explicit form of the terms appearing in these expressions.
The coefficient of the terms can be evaluated by substituting Eq.~(\ref{eq:coeffs}), but here we just want to show 
what kind of processes are generated by $H_\text{eff}^{(1)}$. \\

\underline{Group 2 :  $[g_{ij\sigma},h^\dagger_{i'j'\sigma'}]$}

\begin{enumerate}
\item $i'=i,j'=j:$ $\;\;0$

\item $i'=j,j'=i:$ $\;\;0$

\item $i'=i,j'\neq j$
\begin{enumerate}
\item $\sigma'=\sigma:$$\;\;0$
\item $\sigma'=\bar{\sigma}:$
$\;\;-c^\dagger_{i\sigma}c_{j\sigma}c^\dagger_{i\bar{\sigma}}c_{j'\bar{\sigma}}
\bar{n}_{j\bar{\sigma}}\bar{n}_{j'\sigma}$
\end{enumerate}

\item $i'\neq i,j'=j$
\begin{enumerate}
\item $\sigma'=\sigma:$ $\;\;0$
\item $\sigma'=\bar{\sigma}:$
$\;\;-c^\dagger_{i\sigma}c_{j\sigma}c^\dagger_{i'\bar{\sigma}}c_{j\bar{\sigma}}
 n_{i\bar{\sigma}}n_{i'\sigma}$
\end{enumerate}

\item $i'=j,j'\neq i$
\begin{enumerate}
\item $\sigma'=\sigma:$
$c^\dagger_{i\sigma}c_{j'\sigma}n_{i\bar{\sigma}} n_{j\bar{\sigma}} \bar{n}_{j'\bar{\sigma}}$
\item $\sigma'=\bar{\sigma}:$
$\;\; c^\dagger_{j\bar{\sigma}}c_{j'\bar{\sigma}}c^\dagger_{i\sigma}c_{j\sigma}
 \bar{n}_{j'\sigma} n_{i\bar{\sigma}}$
\end{enumerate}

\item $i'\neq j,j'= i$
\begin{enumerate}
\item $\sigma'=\sigma:$
$-c^\dagger_{i'\sigma}c_{j\sigma}\bar{n}_{i\bar{\sigma}} \bar{n}_{j\bar{\sigma}} n_{i'\bar{\sigma}}$
\item $\sigma'=\bar{\sigma}:$
$c^\dagger_{i'\bar{\sigma}}c_{i\bar{\sigma}}c^\dagger_{i\sigma}c_{j\sigma}
 n_{i'\sigma} \bar{n}_{j\bar{\sigma}}$
\end{enumerate}

\end{enumerate}

\underline{Group 2 :  $[g_{ij\sigma},h_{i'j'\sigma'}]$}\\
\begin{enumerate}
\item $i'=i,j'=j:$ $\;\;0$

\item $i'=j,j'=i:$ $\;\;0$

\item $i'=i,j'\neq j$

\begin{enumerate}
\item $\sigma'=\sigma:\;\;-c^\dagger_{j'\sigma}c_{j\sigma} \bar{n}_{j'\sigma}n_{i\bar{\sigma}} n_{j\sigma}$
\item $\sigma'=\bar{\sigma}:\;\;-c^\dagger_{i\sigma}c_{j\sigma}c^\dagger_{j'\bar{\sigma}}c_{i\bar{\sigma}}n_{j\bar{\sigma}}\bar{n}_{j'\sigma}$
\end{enumerate}

\item $i'\neq i,j'=j$

\begin{enumerate}
\item $\sigma'=\sigma:\;\;c^\dagger_{i\sigma}c_{i'\sigma}\bar{n}_{i\bar{\sigma}}n_{i'\bar{\sigma}}\bar{n}_{j\bar{\sigma}}$
\item $\sigma'=\bar{\sigma}:\;\;-c^\dagger_{i\sigma}c_{j\sigma}c^\dagger_{j\bar{\sigma}}c_{i'\bar{\sigma}}\bar{n}_{i\bar{\sigma}}n_{i'\sigma}$
\end{enumerate}

\item $i'=j,j'\neq i$
\begin{enumerate}
\item $\sigma'=\sigma:\;\;0$
\item $\sigma'=\bar{\sigma}:\;\; c^\dagger_{i\bar{\sigma}}c_{i'\bar{\sigma}}c^\dagger_{i\sigma}c_{j\sigma}n_{i'\sigma}n_{j\sigma}$
\end{enumerate}

\item $i'\neq j,j'= i$

\begin{enumerate}
\item $\sigma'=\sigma:\;\;0$
\item $\sigma'=\bar{\sigma}:\;\; c^\dagger_{j'\bar{\sigma}}c_{j\bar{\sigma}}c^\dagger_{i\sigma}c_{j\sigma}\bar{n}_{j'\sigma}\bar{n}_{i\bar{\sigma}}$
\end{enumerate}

\end{enumerate}

\underline{Group 3 : $[h^\dagger_{ij\sigma},h_{j'i'\sigma'}]$}
\begin{enumerate}
\item $i'=i,j'=j$
\begin{enumerate}
\item $\sigma'=\sigma:$      $\;\;0$
\item $\sigma'=\bar{\sigma}:$    $\;\;c^\dagger_{i\sigma}c_{j\sigma}c^\dagger_{i\bar{\sigma}}c_{j\bar{\sigma}}$
\end{enumerate}

\item $i'=j,j'=i$
\begin{enumerate}
\item $\sigma'=\sigma:$
$\;\;(n_{i\sigma}- n_{j\sigma}) \bar{n}_{j\bar{\sigma}} n_{i\bar{\sigma}}$
\item $\sigma'=\bar{\sigma}:$
$\;\;-c^\dagger_{j\bar{\sigma}}c_{i\bar{\sigma}}c^\dagger_{i\sigma}c_{j\sigma}$
\end{enumerate}

\item $i'=i,j'\neq j$
\begin{enumerate}
\item $\sigma'=\sigma:$ $\;\;0$
\item $\sigma'=\bar{\sigma}:$
$\;\;c^\dagger_{i\sigma}c_{j\sigma}c^\dagger_{i\bar{\sigma}}c_{j'\bar{\sigma}}
\bar{n}_{j\bar{\sigma}}n_{j'\sigma}$
\end{enumerate}

\item $i'\neq i,j'=j$
\begin{enumerate}
\item $\sigma'=\sigma:$ $\;\;0$
\item $\sigma'=\bar{\sigma}:$
$\;\; c^\dagger_{i\sigma}c_{j\sigma}c^\dagger_{i'\bar{\sigma}}c_{j\bar{\sigma}}
 n_{i\bar{\sigma}}\bar{n}_{i'\sigma}$
\end{enumerate}
\item $i'=j,j'\neq i$
\begin{enumerate}
\item $\sigma'=\sigma:$
$\;\;c^\dagger_{i\sigma}c_{j'\sigma}n_{i\bar{\sigma}} \bar{n}_{j\bar{\sigma}} n_{j'\bar{\sigma}}$

\item $\sigma'=\bar{\sigma}:$
$\;\;-c^\dagger_{j\bar{\sigma}}c_{j'\bar{\sigma}}c^\dagger_{i\sigma}c_{j\sigma} n_{j'\sigma} n_{i\bar{\sigma}}$
\end{enumerate}
\item $i'\neq j,j'= i$
\begin{enumerate}
\item $\sigma'=\sigma:$
$\;\;-c^\dagger_{i'\sigma}c_{j\sigma}n_{i\bar{\sigma}} \bar{n}_{j\bar{\sigma}} \bar{n}_{i'\bar{\sigma}}$
\item $\sigma'=\bar{\sigma}:$
$\;\;-c^\dagger_{i'\bar{\sigma}}c_{i\bar{\sigma}}c^\dagger_{i\sigma}c_{j\sigma}
 \bar{n}_{i'\sigma} \bar{n}_{j\bar{\sigma}}$
\end{enumerate}
\end{enumerate}

\noindent\\
{\it B. Spectral function and distribution function in the Floquet prethermal state}\\

In the numerical simulation, we observed a periodic behavior of the time-averaged distribution function in the prethermal states,
$f(\omega+n\Omega)\simeq f(\omega)$.
Here we give an argument why this periodicity appears and how it is connected to the effective temperature 
assuming that the prethermal state is an equilibrium state of the effective Hamiltonian.

In the following we assume the resonant condition, $U=m_0\Omega$. 
We first note that our effective Hamiltonian is for the rotating frame, $H^{\rm rot}(t)=V^\dagger(t) H(t)V(t)$.
Therefore, we also need to transform the expression of the Green's function into the rotating frame. 

First we consider the greater component,
\begin{align}
G^>_{i\sigma,i\sigma}(t,t')&=-i\langle c_{i\sigma}(t) c_{i\sigma}^\dagger(t')\rangle\nonumber\\
&=-i\frac{1}{Z} {\rm Tr} [e^{-\beta H_{\rm eq}}U^{\rm rot}(0,t) V^\dagger (t) c_{i\sigma} V(t) \mathcal{U}^{\rm rot}(t,t') V^\dagger(t') c_{i\sigma}^\dagger V(t')\mathcal{U}^{\rm rot}(t',0)]
\end{align}
Here $H_{\rm eq}=H(0)$, $\mathcal{U}(t,t')=\mathcal{T} \exp\bigl(-i\int^t_{t'} d\bar{t} H(\bar{t})\bigl)=V(t) \mathcal{U}^{\rm rot} (t,t') V^\dagger(t')$ and 
\begin{align}
V^\dagger(t) c_{i\sigma} V(t)=e^{-iU t} c_{i\sigma}n_{i\bar{\sigma}}+c_{i\sigma}\bar{n}_{i\bar{\sigma}}\equiv e^{-iUt} f_{i\sigma}+g_{i\sigma}.
\end{align}
By introducing the time-periodic operators $\tilde{f}(t)\equiv e^{i K_{\rm eff}(t)} f e^{-i K_{\rm eff}(t)}$ and $\tilde{g}(t)\equiv e^{i K_{\rm eff}(t)} g e^{-i K_{\rm eff}(t)}$ and $\tilde{H}_{\rm eq}=e^{iK_{\rm eff}(0)} e^{-\beta H_{\rm eq}} e^{-iK_{\rm eff}(0)}$,
\begin{align}
G^>_{i\sigma,i\sigma}(t,t')=&
-i\frac{1}{Z} {\rm Tr} [e^{-\beta \tilde{H}_{\rm eq}} e^{iH_{\rm eff} t} \tilde{f}_{i\sigma} (t) e^{-i H_{\rm eff} (t-t') }\tilde{f}^\dagger_{\sigma}(t')e^{-iH_{\rm eff}t'}] e^{-iU(t-t')}\nonumber\\
&-i\frac{1}{Z} {\rm Tr} [e^{-\beta \tilde{H}_{\rm eq}} e^{iH_{\rm eff} t} \tilde{g}_{i\sigma} (t) e^{-i H_{\rm eff} (t-t') }\tilde{f}^\dagger_{\sigma}(t')e^{-iH_{\rm eff}t'}] e^{iUt'}\nonumber \\
&-i\frac{1}{Z} {\rm Tr} [e^{-\beta \tilde{H}_{\rm eq}} e^{iH_{\rm eff} t} \tilde{f}_{i\sigma} (t) e^{-i H_{\rm eff} (t-t') }\tilde{g}^\dagger_{\sigma}(t')e^{-iH_{\rm eff}t'}] e^{-iUt}\nonumber\\
&-i\frac{1}{Z} {\rm Tr} [e^{-\beta \tilde{H}_{\rm eq}} e^{iH_{\rm eff} t} \tilde{g}_{i\sigma} (t) e^{-i H_{\rm eff} (t-t') }\tilde{g}^\dagger_{\sigma}(t')e^{-iH_{\rm eff}t'}]. 
\end{align}

Now we make the conjecture that the system reaches a thermal state of $H_{\rm eff}$ with temperature $\beta_{\rm eff}$ after a long enough time. 
In other words, when $t,t'$ are far from the initial time, we can approximate $e^{-\beta \tilde{H}_\text{eq}}/Z$ as $e^{-\beta_{\rm eff}H_{\rm eff}}/Z_{\rm eff}$.
With this assumption and 
\begin{align}
&\tilde{f}(t)=\sum_{n} e^{in\Omega t} \tilde{f}^{(n)},\;\;\tilde{f}^\dagger(t)=\sum_{m} e^{-im\Omega t} \tilde{f}^{(m)\dagger},\nonumber\\
&\tilde{g}(t)=\sum_{n} e^{in\Omega t} \tilde{g}^{(n)},\;\;\tilde{g}^\dagger(t)=\sum_{m} e^{-im\Omega t} \tilde{g}^{(m)\dagger},
\end{align}
we consider the Lehmann representation of the Green's function in terms of eigenstates ($|k\rangle,|l\rangle$) of $H_{\rm eff}$.
The calculation is straitforward and the time averaged Fourier component is 
\begin{align}
\overline{G^>_{i\sigma,i\sigma}}(\omega)&=\int_{-\infty}^\infty dt_r \frac{1}{T} \int_0^T dT_a e^{it_r \omega} G^>_{i\sigma,i\sigma}(t_r;T_a)\nonumber\\
&=-i\frac{2\pi}{Z_{\rm eff}} \sum_{l,k} \sum_m \delta(\omega+\epsilon^{\rm eff}_k-\epsilon^{\rm eff}_l+m\Omega) e^{-\beta_{\rm eff} \epsilon_k^{\rm eff}}
|\langle k| f_{i\sigma}^{m+m_0}+g^m_{i\sigma} |l\rangle|^2.
\end{align}
In the same manner, we can evaluate the lesser Green's function to obtain
\begin{align}
\overline{G^<_{i\sigma,i\sigma}}(\omega)&=\int_{-\infty}^\infty dt_r \frac{1}{T} \int_0^T dT_a e^{it_r \omega} G^<_{i\sigma,i\sigma}(t_r;T_a)\nonumber\\
&=i\frac{2\pi}{Z_{\rm eff}} \sum_{l,k} \sum_m \delta(\omega+\epsilon^{\rm eff}_k-\epsilon^{\rm eff}_l+m\Omega) e^{-\beta_{\rm eff} \epsilon_l^{\rm eff}}
|\langle k| f_{i\sigma}^{m+m_0}+g^m_{i\sigma} |l\rangle|^2.
\end{align}
Now the time-averaged spectral function becomes
\begin{align}
\bar{A^R}(\omega)&=\sum_m \bar{A^R_m}(\omega),\\
\bar{A^R_m}(\omega)&=\frac{1}{Z_{\rm eff}} \sum_{l,k}\delta(\omega+\epsilon^{\rm eff}_k-\epsilon^{\rm eff}_l+m\Omega)( e^{-\beta_{\rm eff} \epsilon_l^{\rm eff}} + e^{-\beta_{\rm eff} \epsilon_k^{\rm eff}})
|\langle k| f_{i\sigma}^{m+m_0}+g^m_{i\sigma} |l\rangle|^2.
\end{align}
Using the spectral function, we can express the lesser Green's function as 
\begin{align}
\overline{G^<_{i\sigma,i\sigma}}(\omega)&=i2\pi \sum_m \bar{A^R_m}(\omega) f(\omega+m\Omega, \beta_{\rm eff}).
\end{align}
Here $ f(\omega+m\Omega, \beta_{\rm eff})$ is the Fermi distribution function at $\beta_{\rm eff}$.
We now further assume that $\bar{A}_m(\omega)$  is finite only around $\omega\in R_m \equiv[-m\Omega-\Delta \epsilon, -m\Omega +\Delta\epsilon]$ with $\Delta \epsilon \ll \Omega$.
In this case the distribution function defined as ${\rm Im} G^<(\omega)/A(\omega)$ is approximately $\frac{1}{e^{\beta_{\rm eff}(\omega+m\Omega)}+1}$ at each $R_m$.
The assumption seems plausible when $H_{\rm eff}$ has small characteristic energy scales compared to $\Omega$.
We note for example that $g^{(0)}\simeq c^\dagger_{\sigma} \bar{n}_{\bar{\sigma}}$ and $f^{(0)}\simeq c^\dagger_{\sigma} n_{\bar{\sigma}}$ and 
these are the leading terms. The rest is higher order corrections that come from the correction to the kick operator.
Therefore, we can expect a similar intensity for $m=0$ and $m=-m_0$ in the spectrum, which should correspond to the 
Hubbard bands.

\end{document}